# Interfacial silica nanoparticles stabilize cocontinuous polymer blends

Lian Bai, John Fruehwirth, Xiang Cheng[*], Christopher W. Macosko[*]

Department of Chemical Engineering and Materials Science, University of Minnesota,

Minneapolis, Minnesota, 55455, USA

**Abstract**

We investigated the formation of cocontinuous structures in polymer blends. These polymeric bijels (bicontinuous interfacially jammed emulsion gels) were composed of polystyrene oligomer, polybutene and fluorescent hydrophobic silica nanoparticles. A micron-sized cocontinuous morphology was stabilized by a monolayer of silica nanoparticles at the interface. Real-time observation of coalescence dynamics in co-continuous polymer blends stabilized by interfacial particles was for the first time achieved via laser scanning confocal microscopy. We demonstrated that suppression of coalescence arises from coverage of interfaces by nanoparticles. Furthermore, by combining confocal microscopy with rheology, we correlated the rheological response of a cocontinuous structure with its morphology change. We found that the rheological behavior can be attributed to competition between interface shrinkage and particle network formation. In addition, we showed that a particle scaffold is maintained even after the remixing of two polymer phases above the spinodal point. Finally, we also discussed differences between the shear response of the particle-stabilized cocontinuous structure and normal colloidal gels: the former one is more fragile than the latter under shear.

## 1. Introduction

Cocontinuous polymer blends are composed of two immiscible or partially miscible polymers in two interpenetrating domains.[1] In contrast to the more accessible droplet-matrix morphology, cocontinuous morphology can impart superior material properties to composites such as enhanced mechanical modulus, impact and electrical conductivity.[1] Moreover, after extracting one phase, cocontinuous polymer blends become porous membranes, which have been used as lithium battery separators[2], supporting substrates for catalysts[3] and scaffolds in tissue engineering[4,5]. Given their importance, it is not surprising that there is growing interest in research of cocontinuous polymer blends [1,6,7].

For two immiscible polymers, cocontinuous blends can be achieved by mechanically melt mixing.[1,7] Partially miscible polymer blends with a miscible region can also form the cocontinuous morphology via the demixing process from spinodal decomposition.[8,9] Regardless the specific route, however, the resulting cocontinuous morphology is intrinsically unstable due to its non-equilibrium nature. Left alone, interfaces between the two polymers will coalesce and the cocontinuous structure will reduce to droplet-matrix morphology.[6,10] To stabilize cocontinuous morphologies, a creative strategy exploiting nanoparticles has been proposed. Nanoparticles trapped at the interface by capillary forces can effectively suppress the coarsening of polymer phases during annealing and stabilize the cocontinuous morphology. For example, Jerome and co-workers added oxidized carbon black (CB) in immiscible polymer blends such as polyethylene/polystyrene (PE/PS)[11] and polystyrene/poly(methyl methacrylate) (PS/PMMA)[12] during melt mixing. Thermodynamic stabilization of cocontinuous morphologies of these polymer blends was achieved thanks to selective localization of CB at the blend interface. Composto *et al.* trapped PMMA grafted silica nanoparticles (SNP) at the interface of a partially miscible polymer blend, poly(methyl methacrylate)/poly(styrene-*co*-acrylonitrile)(PMMA/SAN), which led to cocontinuous morphology after spinodal decomposition with particles jammed at the interface.[13,14,15]

Although a few particle-stabilized cocontinuous polymer blends have been achieved in experiments, to our knowledge no studies have been conducted to systematically investigate dynamics by which particles inhibit the coalescence of cocontinuous polymer structure.[16,13] In contrast, the dynamics of cocontinuous morphology has been recently investigated in the so-called "bi-continuous interfacially jammed emulsion gel" or "bijel". Bijels are composed of two interpenetrating low-viscosity fluids in a cocontinuous structure, which is stabilized by an interfacial colloidal monolayer through spinodal decompostion.[16,3,17] The low viscosity of fluid phases in a bijel enables direct study of the dynamics of interfacial particles during coalescence and the formation of cocontinuous structure.[16,3,17] Here, to exploit the unique feature of bijel systems in the study of polymeric cocontinuous structure, we bridged the separate studies from polymer and colloid fields by introducing a novel polymeric bijel composed of two low molecular weight polymers or oligomers. The system allows us for the first time to investigate the microscopic dynamics of interface coalescence in polymer blends and access the change of particle interfacial coverage in real time.

Specifically, we used a polystyrene oligomer (PS) and a low molecular weight polybutene (PB), which show an UCST phase separation and have a viscosity low enough at room temperature to be suitable for direct imaging. We introduce hydrophobic silica nanoparticles (B-SNP) to stabilize cocontinuous morphology through spinodal decomposition. The new PS/PB/B-SNP bijel system has several unique advantages: First, as already explained, when compared with high molecular weight cocontinuous polymer blends, PS and PB are viscous liquids at room temperature, thus facilitating real time observation of coalescence under confocal microscopy. Second, compared with conventional bijels, the PS/PB/B-SNP system has higher viscosity at room temperature (~10-$10^2$ Pa·s), and its coalescence dynamics are directly relevant to that of high molecular weight cocontinuous polymer blends (Fig. 1). Third, different from bijels where highly polar organic solvents have been used such as water/2,6-lutidine (W/L)[19] and nitromethane/ethylene glycol (NM/EG)[20], PS/PB used in our polymeric bijel, like most high molecular weight polymeric blends formed at high temperature, are much less polar.[21] Such a difference imparts unique particle-matrix and inter-particle interactions, as we show later. Thus, our study can be seen as the first attempt to bridge two separate fields, which may stimulate further investigations on the universal properties of the cocontinuous blend in widely different systems.

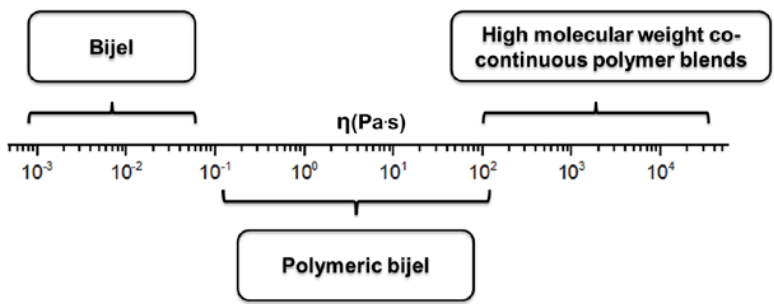

Fig. 1 Viscosity range of bijels, our polymeric bijel and high molecular weight cocontinuous polymer blends.

## 2. Materials and methods

### 2.1 Polymers and particle synthesis

Polystyrene oligomer (PS, Piccolastic A5 Hydrocarbon Resin, $M_n$ = 300, PDI = 1.2, Eastman) and polybutene (PB, PB-24, $M_n$ = 950, Soltex) were used as received. Physical properties of the polymers and the synthesized SNP are listed in Table 1.

Table 1 Material properties

| Materials | Viscosity at 25 °C ($\eta_\infty$, Pa s) | Density ($\rho$, g/cm$^3$) | Refractive index, $n$(20/D) |
|---|---|---|---|
| **PS-A5** | 33.4 | 0.96 | 1.52 ~ 1.55 |
| **PB-24** | 24.1 | 0.89 | ~ 1.49 |
| **SNP** | NA | 2.2 | ~ 1.544 |

We used the well-known Stober method to synthesize monodisperse fluorescent hydrophilic silica nanoparticles (P-SNP).[21,22] Before synthesis, fluorescent cores were prepared by addition of 28 mg of rhodamine B isothiocyanate (RITC, Sigma-Aldrich) to a solution of 5 mL anhydrous ethanol and 44 mg 3-aminopropyltriethoxysilane (APTES, Sigma-Aldrich). The dye solution was then stirred overnight in a 25 mL round-bottom flask at room temperature. The particles were synthesized by adding 7.7 mL ammonia (29% w/w, Sigma-Aldrich), 4.6 mL DI water, 7.7 mL tetraethylorthosilicate (TEOS, > 99%, Sigma-Aldrich) and 5 mL fluorescent core suspension to 176 mL of anhydrous ethanol, and stirring the reaction for 6 hr in a 500 mL round-bottom flask at room temperature. Particle size distribution was characterized by dynamic light scattering (DLS, Brookhaven Instruments Corp. NanoBrook 90Plus Zeta Particle Size Analyzer): the hydrodynamic diameter = 102 nm ± 4 nm. Scanning electron microscopy (SEM, JEOL 6500F) was also used, which confirmed the diameter of the dry particles at ~100 nm.

Hydrophobic modification of the fluorescent P-SNP was achieved via hexamethyldisilazane (HMDS, 98%, Alfa Aesar).[23] An appropriate amount of HMDS was directly added to the reaction mixture of the synthesized P-SNP. The mixture was stirred for 5 days to accomplish the silanization reaction. The fluorescent hydrophobic silica nanoparticles (B-SNP) were subsequently washed twice with ethanol to remove unreacted chemicals and excess fluorescent dye. Finally, the B-SNP was dispersed in tetrahydrofuran (THF) via ultrasonic bath for 10 min before further use during preparation of PS/PB/B-SNP bijels.

**2.2 Cloud point test**

A phase diagram of the PS/PB system was constructed by measuring transmitted light intensity as a function of temperature.[24,25] Samples were loaded in a 1 mL ampoule located in a hole of an aluminum heat stage, allowed to equilibrate in the one-phase region above the UCST of the PS/PB blends and then were cooled to room temperature at a rate of ~ 0.4°C/min. We determined the cloud point when the transmitted light intensity decreased to nearly zero (Fig. 2 inset).

Fig. 2 displays the cloud point curve of neat blends with different weight fractions of PS. Due to kinetic effects, the cloud point is usually located between the binodal and spinodal line.[26] Nevertheless, the cloud point curve reflects the basic phase behavior of the PS/PB used in this study (dash line in Fig. 2): UCST located at ~ 50 °C for 80 wt% PS. Hence, for PS/PB neat blends (50/50 wt%) and the corresponding PS/PB/B-SNP bijels, only a deep and quick quench, such as with dry ice, can move a sample through the metastable region into the unstable region under the spinodal line, where cocontinuous morphology can be formed via spinodal decomposition.[27]

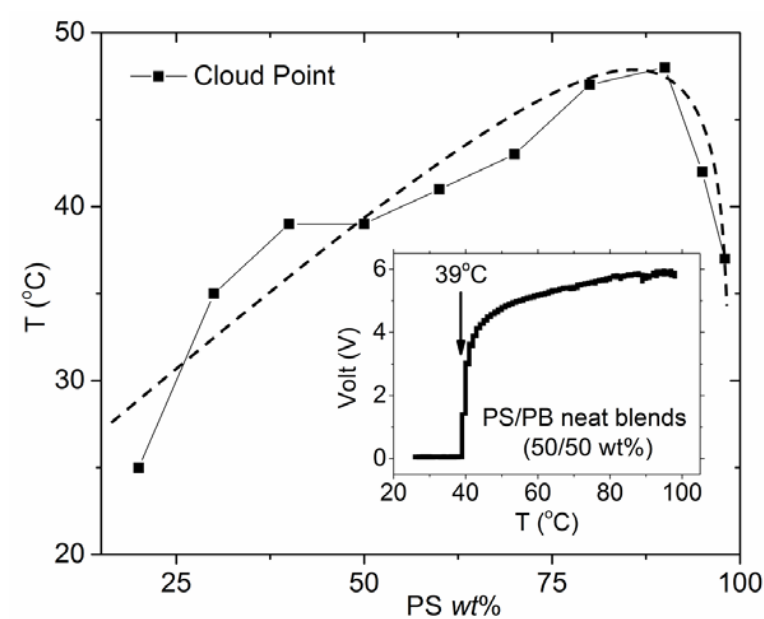

Fig. 2 Cloud point curve of PS/PB neat blends with different PS composition. The dashed line was drawn to guide the eye. The inset illustrates determination of the cloud point for a 50/50 wt% neat blend. Transmitted light intensity (photocell voltage) is plotted as a function of temperature.

**2.3 Polymeric bijels preparation**

Equal amounts of PS and PB were dissolved in a suspension of B-SNP in THF. The mixture was dried in an oven at 80 °C for 24 hr to remove the THF. The PS/PB/SNP mixtures were denoted by ($A/B/x$), where $A$ and $B$ represent the weight fraction of PS and PB in the binary blend, respectively, and $x$ is the weight fraction of SNP with respect to the total amount of polymer. In this study, $x$ = 0.5, 2.0 and 4.5 were used. Since the UCST-type phase-separation temperature is ~ 40 °C for PS/PB (50/50 wt%) blends (Fig. 2), the resulting PS/PB/SNP mixtures are homogenous at 80 °C. The PS/PB/SNP (50/50/$x$) mixture was transferred to a sample cell (Fig. 3a) which was preheated to 80 °C. The cell was quenched below the phase-separation temperature by placing it between dry ice blocks (-78 °C) causing the cocontinuous morphology to form through spinodal decomposition.

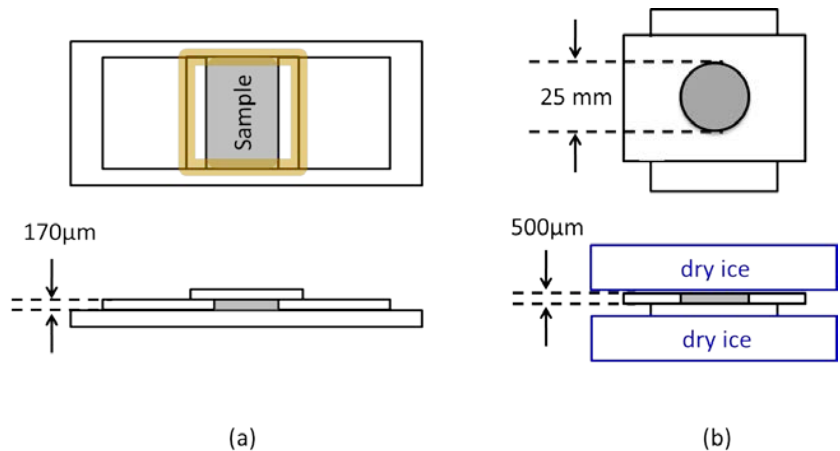

Fig. 3 (a) Top view and cross section of the confocal sample cell: a bottom glass side serves as the holder, two pieces of cover glass in the middle as spacers and a cover glass on the top for microscope observation. Epoxy (yellow) was used to glue the cover glasses on the bottom slide. The sample is contained in the gray region. (b) Top view and cross section of the mold for preparing polymeric bijels for rheological measurements: a bottom stainless steel plate topped with another stainless steel plate with a central hole 25 mm in diameter. The sample is placed inside the hole (gray area). Dry ice in a sandwich structure is used to quench the sample into a cocontinuous morphology.

### 2.4 Confocal microscopy

We used laser scanning confocal microscopy (LSCM, Olympus Fluo View 1000) to study the dynamics of the PS/PB/B-SNP bijels. Unless explicitly stated, the images were always taken at room temperature (20 °C).

The refractive index difference between the two polymers and SNP is small enough to allow us to image the sample ~30 μm below the coverslip. We used two different laser channels for imaging. Picolastic PS was fluorescent when excited by a 488 nm laser beam, allowing the two polymer phases to be distinguished in the first channel (green region in Fig. 4). A second, 563 nm laser was used to excite the RITC dye in the SNP, allowing detection of SNP through the second channel (red dots in Fig. 4). Note that due to the small size of the particles, we cannot identify individual SNP in our images. The locations of particles were determined within ~ 200 nm, a limit set by the diffraction of light.

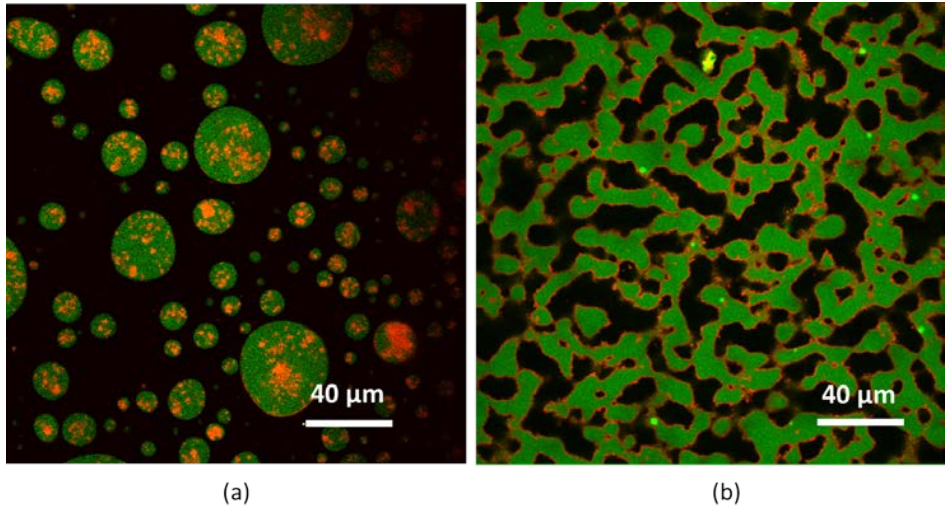

Fig. 4 Confocal images of (a) PS/PB/P-SNP (50/50/4.5) droplet-matrix; (b) PS/PB/B-SNP (50/50/4.5) bijels. The green region is PS, black region is PB and the red dots represent particles.

To analyze the domain size change during coalescence, we thresholded the confocal images to create binary images. The interface length, $L_{int}$, was then obtained by tracking the interface between the black and white regions in the binary images. The characteristic domain size ($\xi$) can thus be defined as $1/Q$, where $Q$ is the interfacial length per unit area, $Q = L_{int.}/S$ and $S$ is the total area of the confocal image.

Similar binary images can also been obtained from the second channel with the 563 nm laser and the location of SNP estimated. The confocal images from these two channels can be further combined to yield the degree of interfacial coverage by nanoparticles. In this analysis, we directly compare the position of the polymer-polymer interface from the first channel with the position of nanoparticles from the second channel. Interfacial coverage is defined as the fraction of the polymer-polymer interface that is occupied by nanoparticles. Due to the pixel noise of imaging, the following criterion was adopted to determine coverage: the polymer-polymer interface from the first channel is covered by particles only if the corresponding pixel or one of the eight nearest neighboring pixels around the corresponding pixel from the second channel is occupied by particles.

## 2.5 Cryogenic scanning electron microscopy

We also used cryogenic scanning electron microscopy (cryo-SEM) to visualize the location of B-SNP. Cryo-SEM images were taken using a Hitachi S4700 cold field emission gun scanning electron microscope. In the first step, the PS/PB/B-SNP bijels were placed in a small capillary then quenched in liquid nitrogen. Afterwards, we transferred the capillary to a preparation chamber (-160 $^{o}$C), where the sample was fractured then coated with a thin platinum layer. Finally, the sample was transferred to the cryo-SEM chamber (-130 $^{o}$C) under high vacuum for SEM imaging.

## 2.6 Rheological characterization

Rheological measurements of the PS/PB neat blends and the PS/PB/B-SNP bijels were performed on a rotational rheometer (AR-G2, TA Instruments) with a Peltier temperature control stage as the bottom plate. We used the parallel plate geometry with a diameter of 25 mm and a gap height of 400 μm. Polymeric bijels were formed by quenching from 80 $^{o}$C to -78 ºC in a sandwich structure by dry ice blocks (Fig. 3b). Afterwards, we removed the spacer and transferred the disk-shaped bijel sample on the plate holder to the Peltier stage (0 $^{o}$C) of the rheometer. After setting the gap and stabilizing the plate holder on the stage with conductive tape, the sample was heated back to room temperature (20 $^{o}$C) to take a dynamic time sweep at 0.1% strain and 1 rad/s. In addition, we confirmed that small angle oscillatory shear (SAOS) from the dynamic time sweep does not change the bijels' morphology at room temperature. Frequency sweeps were also performed from 0.02 to 100 rad/s.

## 3. Results and discussion

### 3.1 Formation and microstructure of PS/PB/B-SNP bijels

Even though cocontinuous morphology can be achieved for the neat PS/PB blends via spinodal decomposition by a deep quench the cocontinuous structure quickly coalesces and collapses to droplet-matrix morphology. To stabilize the cocontinuous morphology, we tested silica nanoparticles with different wetting properties. For hydrophilic silica nanoparticles (P-SNP), the PS/PB/P-SNP (50/50/4.5) mixtures gave droplet-matrix morphology (Fig. 4a). The particles aggregated in the more polar PS phase. In contrast, cocontinuous morphology was achieved when the hydrophobic silica nanoparticles (B-SNP) were used (Fig. 4b and supplementary information, Video SV1). The cryo-SEM images in Fig. 5 show that the particles are located at the interface between the two phases and, moreover, that they are lined up in monolayers.

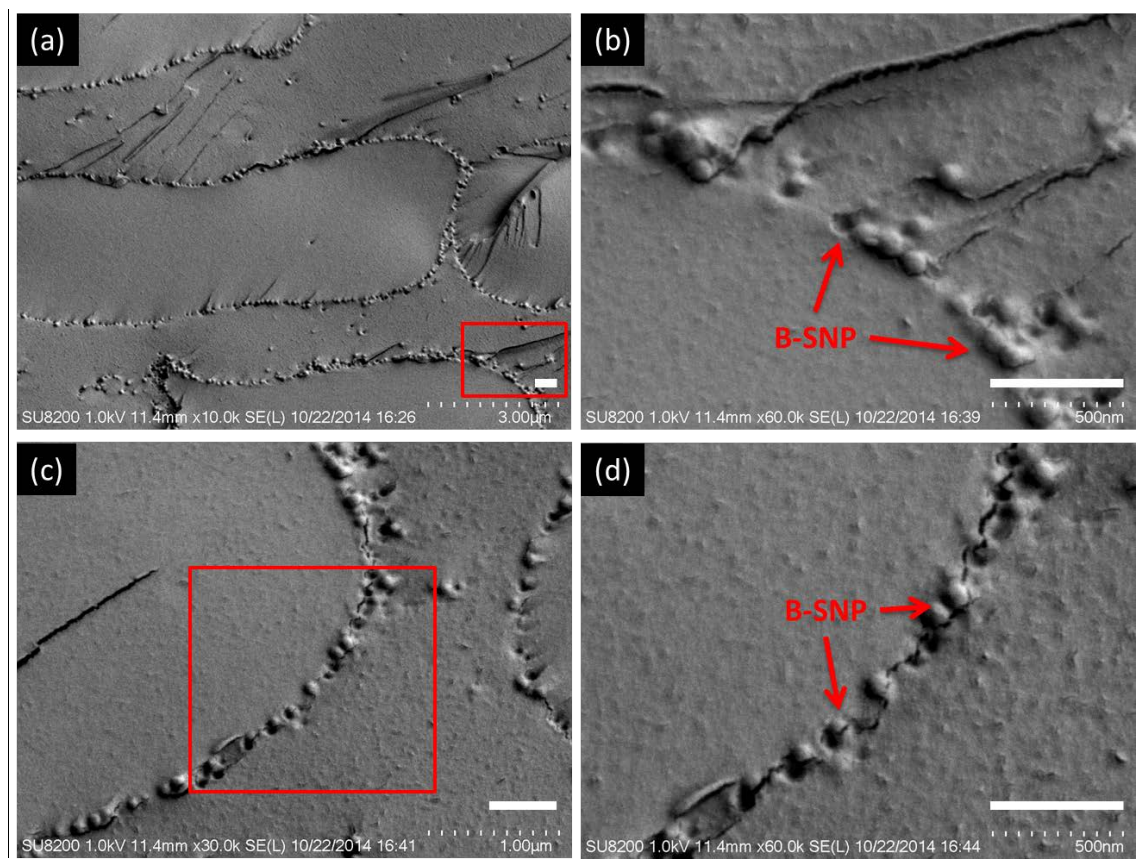

Fig. 5 Cryo-SEM images of the PS/PB/B-SNP (50/50/4.5) bijels. (b) and (d) are magnified views of the rectangular regions indicated in (a) and (c). Particles are indicated by arrows. The scale bar is 500 nm.

Trapping of hydrophobic particles at the interface can be easily understood from a thermodynamic perspective. According to Young's equation,[29,30] the wettability coefficient $\omega$ is defined as,

$$\omega = \cos\theta = (\gamma_{\text{Si-PB}} - \gamma_{\text{Si-PB}}) / \gamma_{\text{PS-PB}} \qquad (1)$$

where $\theta$ is the particle contact angle at the interface (Fig. 6), $\gamma_{\text{PS-PB}}$ is interfacial tension between PS and PB, and $\gamma_{\text{Si-PB}}$ and $\gamma_{\text{Si-PB}}$ are the interfacial tensions between the particle and PB and PS, respectively. The criterion for locating particles at the interface is simply $-1 < \omega < 1$ ($0° < \theta < 180°$); in contrast, particles are located in the PB phase when $\omega > 1$ and in the PS phase when $\omega < -1$. We use the Owens-Wendt-Rabel-Kaelble (WORK) approximation to estimate the interfacial energies.[30]

$$\gamma_{\text{PS-PB}} = \gamma_{\text{PS}} + \gamma_{\text{PB}} - 2\sqrt{\gamma^d_{PS}\gamma^d_{PB}} - 2\sqrt{\gamma^p_{PS}\gamma^p_{PB}} \qquad (2)$$

where $\gamma^d_i$ and $\gamma^p_i$ are the dispersive and polar parts of the surface tension of component $i$ in the system, which are available in the literature for PB, PS, P-SNP and B-SNP (Table 2).[31,32] From Eq. (1) and (2), the wetting coefficients for PS/PB/P-SNP and PS/PB/B-SNP systems are −4.65 and 0.12, respectively. These results agree with the experimental observation, i.e., P-SNP aggregate in the PS phase while B-SNP locate at the interface.

Table 2 Surface tensions and wetting coefficients of the blends

| System | Interfacial tension (mN/m)* | | $\omega$ | Particles location |
|---|---|---|---|---|
| **PS / PB /P-SNP** | PS / PB | 6.21 | | |
| | PS / P-SNP | 21.67 | −4.65 | PS phase |
| | PB / P-SNP | 50.74 | | |
| **PS / PB /B-SNP** | PS/PB | 6.21 | | |
| | PS / B-SNP | 1.37 | 0.12 | interface |
| | PB / B-SNP | 2.10 | | |

* The surface tensions of each component are (in mN/m): $\gamma_{\text{PS}} = 40.7$ ($\gamma^d_{PS} = 34.5$, $\gamma^p_{PS} = 6.1$), $\gamma_{\text{PB}} = 33.6$ ($\gamma^d_{PB} = 33.6$, $\gamma^p_{PB} = 0$), $\gamma_{\text{P-SNP}} = 80.0$ ($\gamma^d_{P-SNP} = 29.4$, $\gamma^p_{P-SNP} = 50.6$), $\gamma_{\text{B-SNP}} = 32.0$ ($\gamma^d_{B-SNP} = 30.0$, $\gamma^p_{B-SNP} = 2.0$).

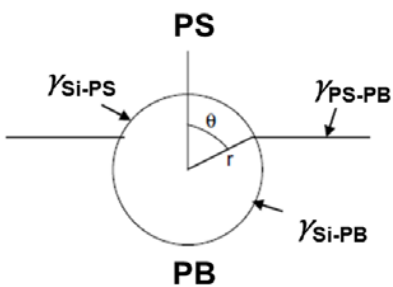

Fig. 6 Schematic of a spherical silica nanoparticle at the interface between PS and PB.

## 3.2 Dynamics of coalescence

Fig. 7a shows the characteristic domain size ($\xi$) plotted against annealing time at room temperature for both the neat blend and PS/PB/B-SNP bijels with different particle concentrations (Supplementary Video SV2 – S5). As can be seen from the inset of Fig. 7a, the neat blend shows a continuous increase in characteristic size. The change of morphology for the neat blend is shown in Fig. 8a and Video SV2 in the Supplementary Information, which displays the change from an initial cocontinuous structure with small domain size to a final drop-matrix structure.

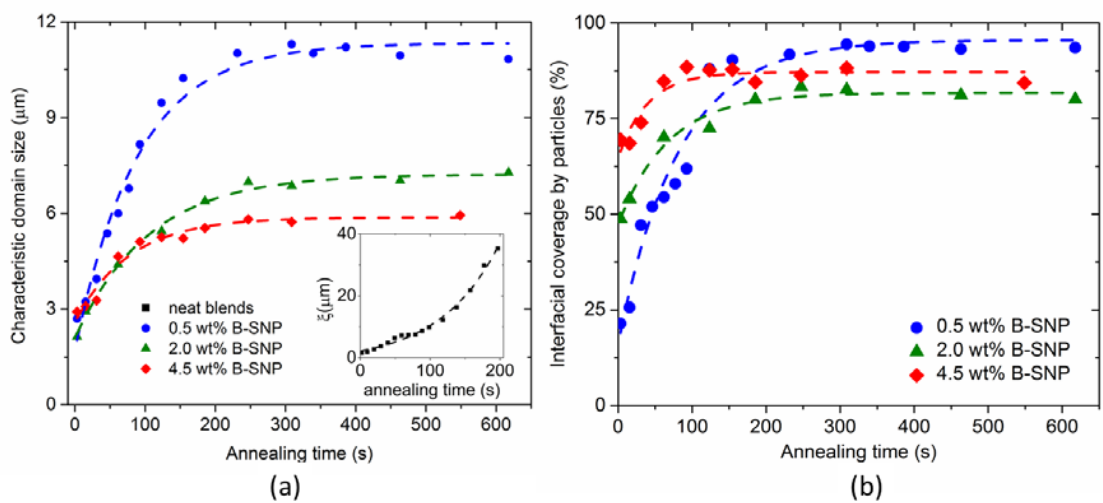

Fig. 7 (a) Characteristic domain size ($\xi$) as a function of annealing time for PS/PB neat blends (inset) and PS/PB/B-SNP bijels with different weight fraction of particles at 20 $^{o}$C ; (b) Interfacial coverage by particles as a function of annealing time for PS/PB/B-SNP bijels with different weight fraction of particles at 20 $^{o}$C. The dashed lines in both figures are exponential fits (Table 3).

In contrast, the characteristic domain size of PS/PB/B-SNP (50/50/0.5) bijels increases initially then plateaus at later times (Fig. 7a). With low particle concentration (0.5 wt%) the cocontinuous structure is stabilized with a large domain size (Fig. 8b and Video SV3 in the Supplementary Information). Similar trends were observed when the weight fraction of particles is increased to 2.0 wt% and 4.5wt% (Fig. 7a). In these cases, the cocontinuous morphology changes much less during annealing and stabilizes at a smaller domain size (Fig. 8c and Video SV4 and SV5 in the Supplementary Information), indicating that coalescence of the cocontinuous morphology is more effectively suppressed.

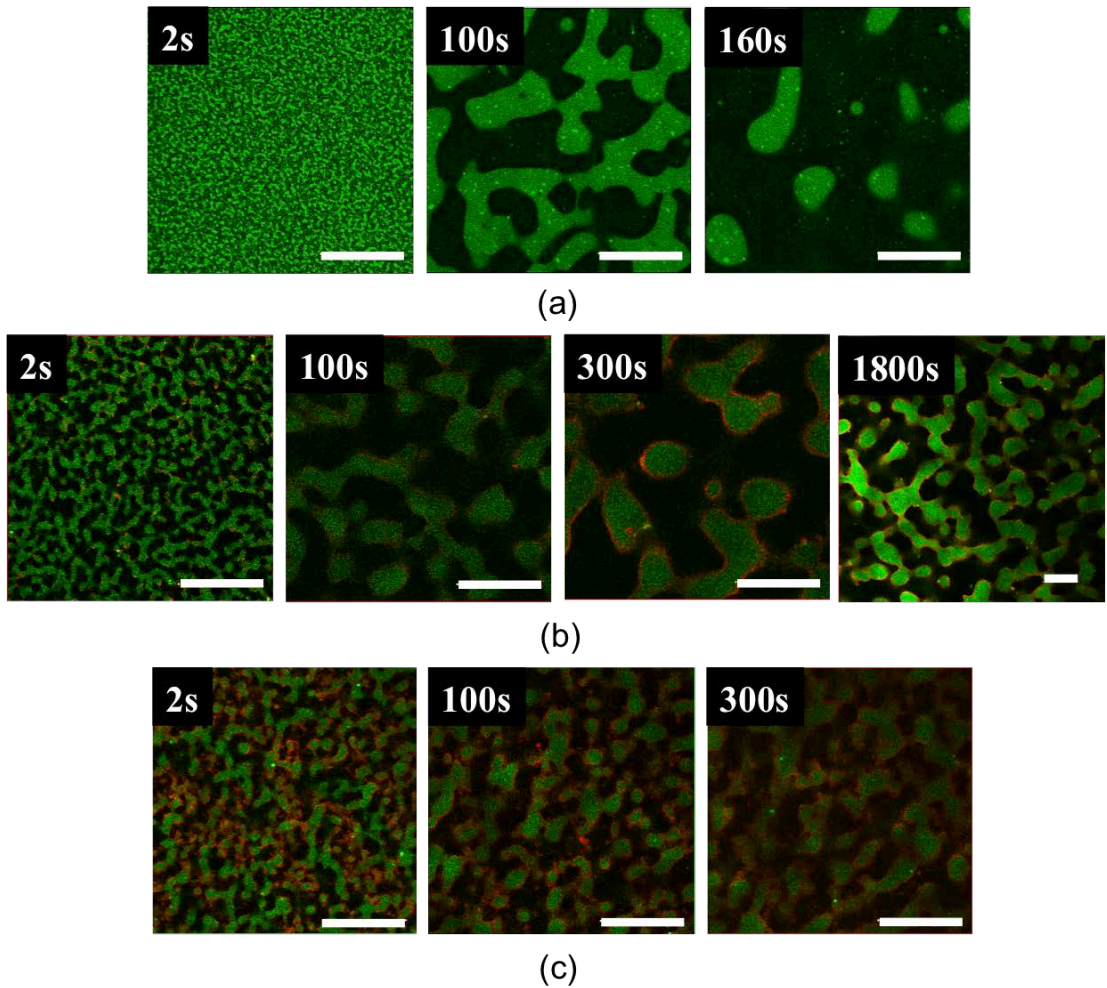

Fig. 8 Confocal images from real time observation of (a) PS/PB neat blends, (b) PS/PB/B-SNP (50/50/0.5) bijels and (c) PS/PB/B-SNP (50/50/4.5) bijels. The scale bar is 40 μm.

To unambiguously demonstrate the effect of particles in stabilizing the cocontinuous structure, we plotted interfacial coverage by particles (SNP %) as a function of annealing time in Fig. 7b. The B-SNP (red dots in Fig. 8b and 8c when $t = 2$ s) are sparsely distributed on the interface of the blends at the beginning of annealing. The initial coverage depends on initial particle concentration (Fig. 7b). As the interfacial area shrinks with annealing, coverage increases. The correlation between the morphology of the blends and interfacial particle coverage is most pronounced when comparing Fig. 7a and 7b: the increase of interfacial coverage is accompanied by a slow-down of coalescence. Interfacial coverage reaches a plateau around the time when domain size also reaches a constant value. The plateau value in interfacial coverage is slightly smaller than 100%, presumably due to the finite resolution and pixel noises of our imaging.

Quantitatively, we fit the dynamics of domain size ($\xi$) and interfacial particle coverage (SNP %) using a simple exponential form:

$$y = y_0 + Ae^{-(t-t_0)/\tau} \qquad (3)$$

where $y$ can be $\xi$ or SNP %, $y_0$ is their plateau value, $A$, $t_0$ and $\tau$ are fitting parameters where $\tau$ is the relaxation time of the process. As shown in Table 3, the domain size and the interfacial particle coverage show similar relaxation times illustrating the correlation between interfacial coverage and stabilization of cocontinuous morphology. Furthermore, the results also show that a higher particle concentration leads to a faster suppression of the coalescence.

Table 3 Characteristic times from exponential fitting of characteristic domains size, interfacial coverage by particles and G'

| PS/PB/B-SNP bijels (50/50/x) | Characteristic time (s) | | |
|---|---|---|---|
| | from domain size ($\tau$ in Eq. 3) | from interfacial coverage ($\tau$ in Eq. 3) | from $G'$ time sweep (minimum valley values of $G'$) |
| 0.5 wt% | 89.2 ± 9.2 | 81.5 ± 11.3 | 143.5 ± 15.0 |
| 2.0 wt% | 105.1 ± 9.8 | 67.5 ± 13.2 | 104.0 ± 15.0 |
| 4.5 wt% | 76.8 ± 13.6 | 39.8 ± 14.1 | 26.0 ± 15.0 |

When particles cover the interface (plateau of SNP % in Fig. 7b), they suppress and even prohibit coalescence in two ways. First, interfacial particles increase the rigidity of the cocontinuous interface. The increase in rigidity decreases deformability of the interface, thus retarding shrinkage of interfacial area. When the interface is full of particles, it cannot further shrink and coarsening of the interface stops. Second, the interfacial particles form a solid barrier preventing the coalescence of contacting domains from the same phase—a mechanism that also prevents the coalescence of Pickering emulsions.[16]

### 3.3 Rheology of PS/PB/B-SNP bijels

Confocal microscopy illustrated the dynamics of morphology change during coalescence, and the rheological properties also changed with morphology. We measured the storage ($G'$) and loss ($G''$) moduli of PS/PB/B-SNP bijels during coalescence at room temperature and compared the rheology with the dynamics of cocontinuous structure from direct confocal imaging. Three different mechanisms contribute to the storage modulus ($G'$) of PS/PB/B-SNP bijels: the elasticity of the polymer components ($G'_{comp}$), the capillary restoration of the interfaces ($G'_{int}$) and the particle network at the interface ($G'_{SNP}$).[7] Since $G'_{comp}$ from the pure PS and PB components are small (~ 0.1-1Pa), we can safely neglect its contribution. Therefore, $G' \approx G'_{int} + G'_{SNP}$.

The time evolution of $G'$ of PS/PB neat blends and PS/PB/B-SNP bijels during annealing at room temperature is shown in Fig. 9. The rheological behavior of PS/PB (50/50 wt%) neat blends during the time sweep is similar for many other cocontinuous polymer blends during annealing[7,33], namely, an apparent initial decrease of $G'$ followed by a gradual decay with slow rate to reach a plateau at long times. The decrease of elasticity is attributed to the shrinkage of interfacial area and finally the breakup of the cocontinuous structure into droplets.

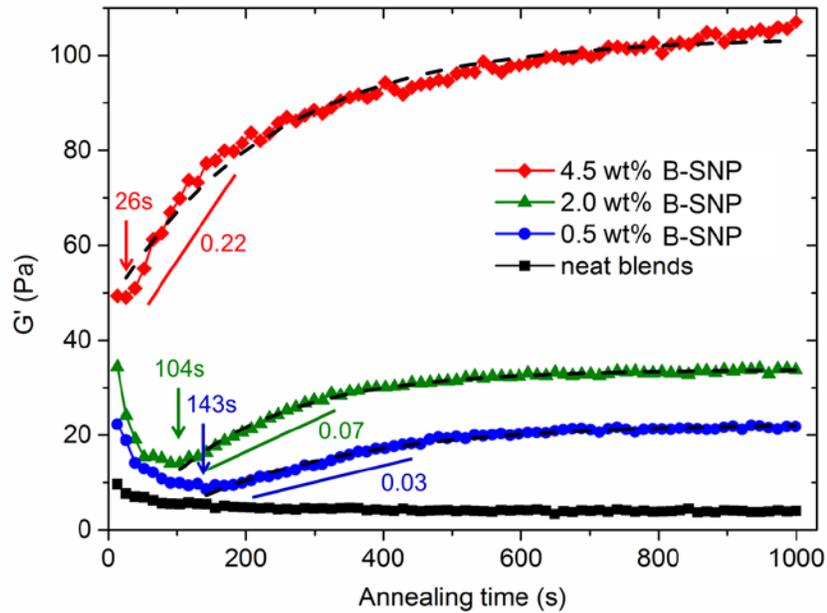

Fig. 9 Elastic moduli ($G'$) as a function of annealing time for PS/PB (50/50 wt%) neat blends and PS/PB/B-SNP bijels with different weight fractions of B-SNP. All data were measured at 20 °C, 1 rad/s and 0.1% strain. The dashed lines are exponential fits. The times for the minimum values of $G'$ and the rate of increase in G' with annealing time are also indicated.

The rheological behavior of PS/PB/B-SNP bijels is quite different from that of neat blends: a rapid $G'$ decrease in the beginning is followed by a large increase, which eventually plateaus with a plateau value much larger than that of neat blends. The variation of $G'$ with the annealing time directly reflects the underlying structural changes in the bijels and can thus be qualitatively understood based on the competition of two opposite structure-induced rheological changes. As shown in Fig. 7 and Fig. 8, during coalescence, interfacial area shrinks with the increase of domain size, leading to the decrease of $G'_{int}$. Meanwhile, a particle network forms at the interface, resulting in an increase of $G'_{SNP}$. For the PS/PB/B-SNP (50/50/4.5) bijels, the minimum of $G'$ occurs at 26 s followed by a steep increase (slope = 0.22 Pa/s). When particle loading is reduced, the minimum of $G'$ appears at a later time (104 s for 2.0 wt% B-SNP and 143 s for 0.5 wt% B-SNP). The increase of $G'$ also becomes much slower (slope = 0.07 Pa/s for 2.0 wt% B-SNP and 0.03 Pa/s for 0.5 wt% B-SNP). Therefore, the effect of $G'_{SNP}$ gradually decreases with the decrease of particle loading. The annealing time at the minimal $G'$ characterizes the structural change of co-continuous phases, which quantitatively matches the time scale measured from direct imaging (Table 3). In addition, bijels with higher particle loading and smaller domain size also have larger elastic modulus ($G'$ plateau value in Fig. 9). Similar results were also observed in other bijel systems.[19,20]

After reaching the minimum, $G'$ slowly increases and takes much longer time (~ 500 s for PS/PB/B-SNP (50/50/0.5) bijels) to reach the plateau. This rheological feature indicates that particles still rearrange along the polymer-polymer interface long after the morphology of the interface has been stabilized. The rearrangement leads to denser particle packing and thus larger $G'$. Due to the resolution limit of confocal microscopy, the rearrangement of particles at the interface cannot be directly observed. Note that for PS/PB/B-SNP (50/50/4.5) bijels, the plateau of $G'$ cannot even be reached within our measurement window (> 1000 s). Such a slow increase may be explained by slower particle rearrangement on the interface with higher interfacial curvature in PS/PB/B-SNP (50/50/4.5) bijels. It has been shown that a larger spatial curvature leads to more frustrated particle packing and, therefore, slower particle dynamics in the glassy state.[33]

Finally, we also performed dynamic frequency sweeps of different PS/PB/B-SNP bijels after the time sweep when the morphology has been stabilized by interfacial particles (Fig. 10a, 10b and 10c). Unlike conventional bijel systems where $G'$ exceeds $G''$ when the bijel forms after spinodal decomposition,[24,21] $G''$ of the PS/PB/B-SNP polymeric bijels is larger than their $G'$. The high value of $G''$ is due to the high viscosity of the PS/PB polymer matrix in polymeric bijel, as compared with the low viscosity organic solvents in conventional bijels.

While the blend interface has no effect on the value of $G''$, the trend of $G'$ at low frequencies provides important information on the morphology of the blends. We plot $G'$ of different PS/PB/B-SNP bijels as a function of frequency in Fig. 10d. All three polymeric bijels show a power-law relation at low frequencies, $G' \propto \omega^n$ with $n = 0.37$, 0.62 and 0.84 for PS/PB/B-SNP bijels with 4.5 wt%, 2.0 wt% and 0.5 wt% B-SNP, respectively. The power-law relation has been suggested as a rheological signature of cocontinuous morphology.[33,7,9] As confocal microscopy can only provide morphology information within ~30 μm from the sample surface, the rheological measurements here provide a sensitive supplement to confirm the formation of cocontinuous structure throughout the whole samples.

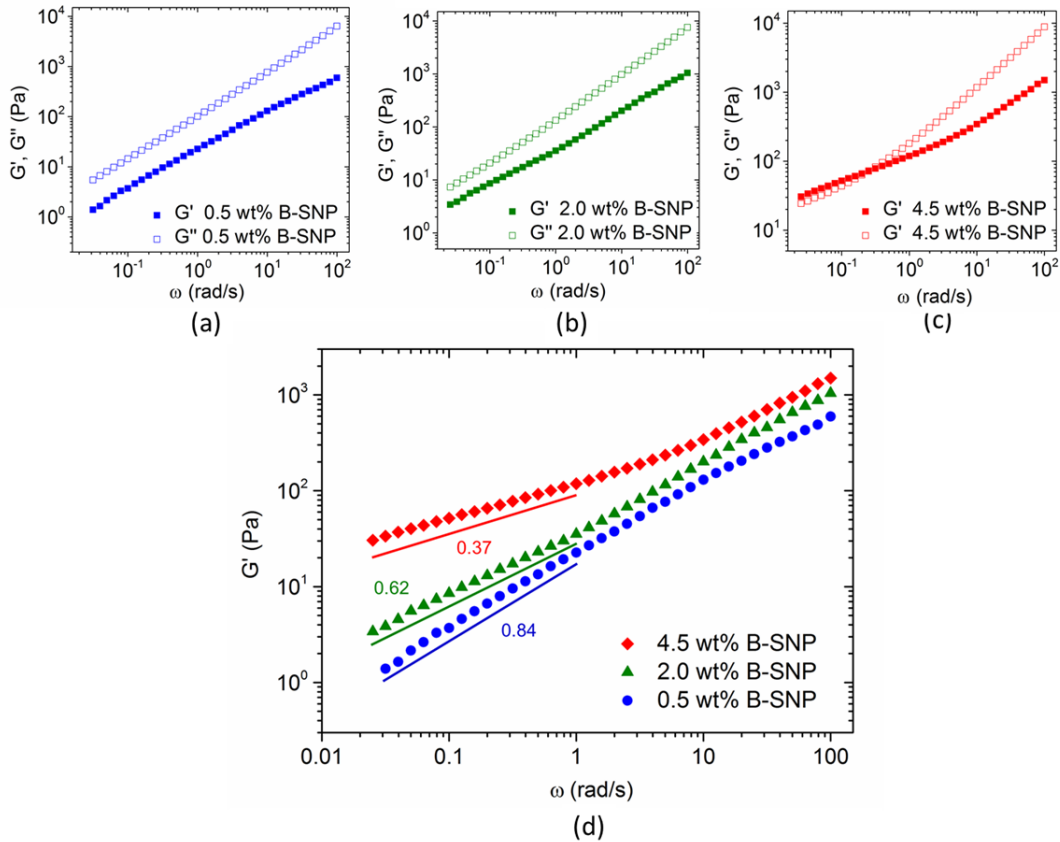

Fig. 10 Dynamic frequency sweeps of PS/PB/B-SNP bijels with different particle loading: (a) 0.5 wt% B-SNP; (b) 2.0 wt% B-SNP; (c) 4.5 wt% B-SNP. (d) The dynamic elastic moduli ($G'$) as a function of frequency for PS/PB/B-SNP bijels with different weight fractions of B-SNP. Measured from low to high frequency at 20$^{\circ}$C and 0.1 % strain. Slopes at low frequencies are also indicated.

### 3.4 Formation of monogel

When the PS/PB/B-SNP bijels were reheated from room temperature back to 80$^{\circ}$C, PS and PB remixed into a single liquid phase. However, the jammed particle monolayer still remains intact where the interface existed before remixing (Fig. 11a). This particle network or "skeleton" structure is called a "monogel" because it comprises a web of locally planar arrangement of particle monolayers.[34] When the particle monogel was studied, confocal images were taken at 80 °C with a temperature controlled heating stage to maintain the temperature of the whole sample cell.

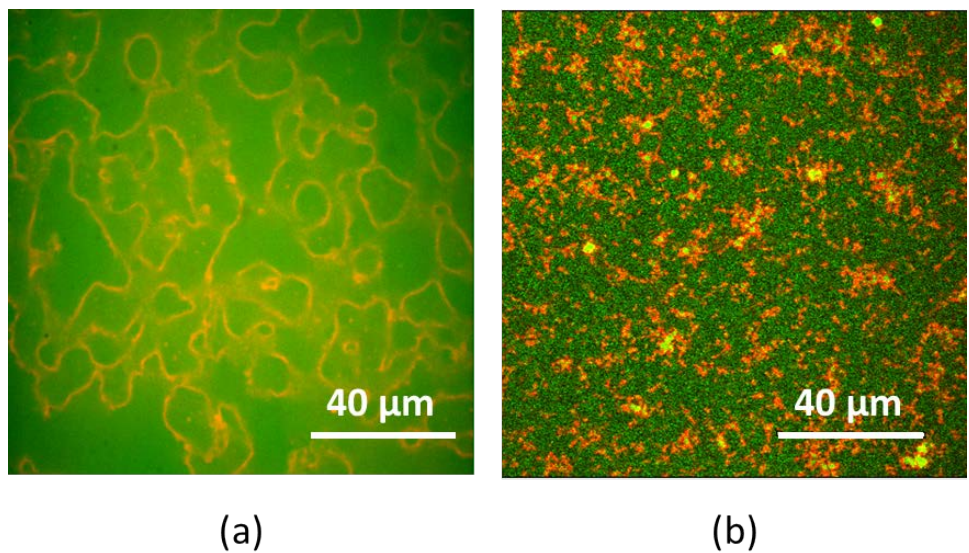

(a)  (b)

Fig. 11 Confocal images of the monogel from PS/PB/B-SNP polymeric bijel at 80 °C: (a) before shear; (b) after shear. The green region is the remixed single-phase polymer matrix, and the red dots represent the hydrophobic particles.

Monogel has also been observed in bijels composed of water and 2,6-lutidine (W/L)[34,20], where the formation of monogel has been attributed to the stabilization of particles in the primary minimum of the DLVO (Derjaguin-Landau-Verweg-Overbeek) potential[35]. Attractive capillary forces and interfacial jamming in the W/L systems overcome electrostatic repulsion and drive particles across the energy barrier of the DLVO potential into the primary van der Waals minimum. In the W/L systems, short-ranged attraction stems from the van der Waals force between particles while long-ranged repulsion is due to the weak, negative electrostatic charge of the dissociated silanol groups on the surface of particles.[36,34] In our PS/PB/B-SNP polymeric bijels, we use hydrophobic particles with HMDS graft layers for stabilization. In conventional bijels (e.g. nitromethane and ethylene glycol (NM/EG) with B-SNP), HMDS graft layers on B-SNP hinder the formation of monogel.[20] This is due to the solvation of HMDS layers by the dissociated nitromethane molecules, which enhances the negative electrostatic charges on the particles. Therefore, the monogel in NM/EG system cannot form because long-range repulsion overcomes short-range attraction and impedes rearrangement of particles.[20] However, in the polymeric bijel, both PS and PB molecules are hardly dissociated and thus do not solvate the HMDS graft layers. It is highly possible that the particles in the PS/PB polymeric bijel experience a smaller repulsive barrier compared with that of W/L bijel and are easily driven to the primary minimum by capillary attractions. Therefore, the less polar polymer molecules in the polymeric bijel facilitate the formation of monogel with hydrophobic particles.

Monogel can be treated as a low-density colloidal gel in a single-phase fluid, thus we compare the rheological behavior of the monogel formed in the polymeric bijel with conventional colloidal gels. Large amplitude oscillatory shear (LAOS) sweeps were run at 80$^{o}$C (Fig. 12). As shown in the first LAOS sweep, the monogel displays signatures of gel-like rheology, with solid-like viscoelasticity at small γ, ($G_o$' > $G_o$''), and a crossover of $G$' and $G$'' at strain of around 1.0 %, which is often regarded as the onset of yielding. This rheological behavior under LAOS is also observed in many colloidal gel systems. [37,38,39]

However, the rheological behavior of the monogel in the second LAOS sweep after the first LAOS is quite different: $G_o$' decreased by an order of magnitude compared with the first LASO sweep and $G$' is always smaller than $G$''. The differences between these two LAOS sweeps can be explained by the morphology change of the monogel sample due to shear during the first LAOS sweep (Fig. 11b). The particle network is completely destroyed by shear and collapses into clusters of particles dispersed in the single phase matrix. The process is irreversible. Without the process of spinodal decomposition serving as a template to guide particles, the dispersed particles cannot re-aggregate in the form of locally planar monolayers. This behavior of monogel under LAOS is different from that of conventional colloidal gels, which usually return to a gelled state when shear ceases.[20] The monogel is a more fragile state.

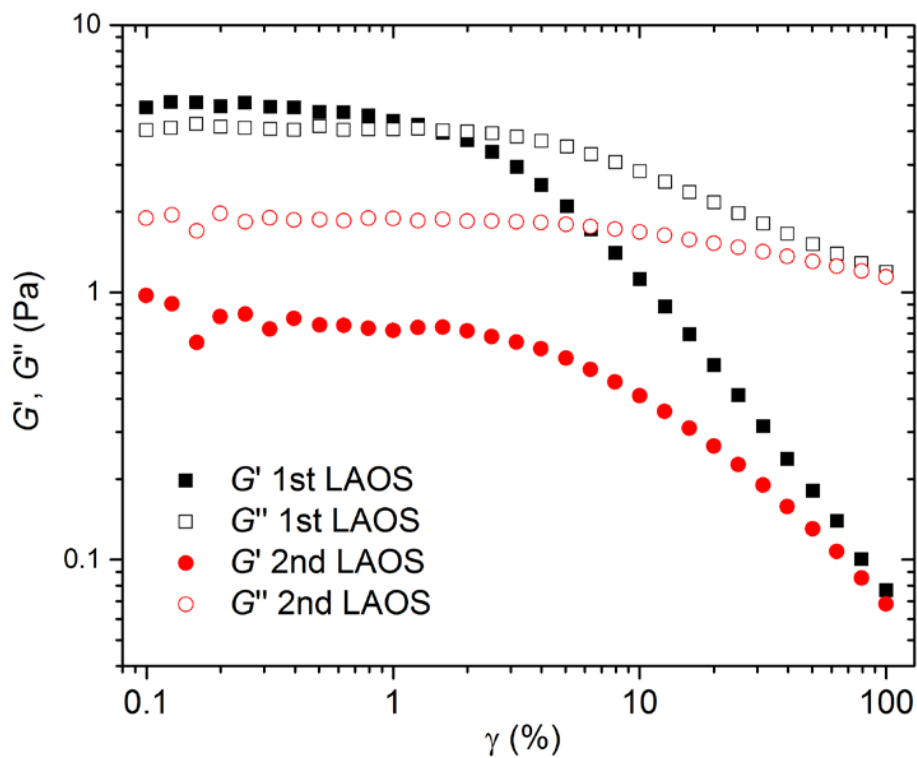

Fig. 12 The storage (*G'*) and loss (*G"*) moduli of the monogel as a function of strain (γ) in twice LAOS sweeps. All data were measured at 80°C with a frequency of 1 rad/s.

## 4. Conclusions

In summary, we developed a polymeric bijel system consisting of hydrophobic silica nanoparticles (B-SNP) and two low molecular weight polymers, polystyrene (PS) and polybutene (PB). We used these bijels to study the mechanism by which interfacial silica nanoparticles suppress the coalescence of cocontinuous polymer blends. Interfacial location of the silica was demonstrated both by confocal microscopy and high magnification cryo-scanning electron microscopy. Real-time dynamics of coalescence of the PS/PB/B-SNP bijels with different particle loadings were investigated via confocal microscopy. We demonstrated that interfacial coverage by particles leads to suppression of coalescence and gives rise to stabilization of cocontinuous structure. In addition, rheological measurements were taken to correlate rheology with morphology.

Competition between the shrinkage of interface and the formation of a particle network results in non-monotonic rheological response of PS/PB/B-SNP bijels during annealing. Our measurements also indicate the existence of shear-induced particle rearrangement along the interface after the morphology has been stabilized. Finally, monogels form by heating the PS/PB/B-SNP polymeric bijels back to 80 $^{o}$C and remixing the PS and PB to one single fluid phase. We observed for the first time that hydrophobic particles form a monogel from direct remixing. Moreover, the relationship between monogel and conventional colloidal gels was studied via LAOS sweeps and confocal microscopy. Although monogels and colloidal gels show similar rheological response during shear, monogels cannot reform after shear. One possible direction for future research is to develop a confocal rheometer with temperature control that can simultaneously measure the rheology with the morphology change in both the bijel and the monogel phases.

## 5. Acknowledgement

The authors are grateful to Dr. Garrett Swindlehurst for his help in the synthesis of the fluorescent silica nanoparticles, Dr. Guillermo Marques and Dr. John Oja of University Imaging Centers for their help in image acquisition, Dr. Aaron Hedegaard and Miss Yanlin Liu for their help in image processing and analysis. The authors thank the members of the University of Minnesota Industrial Partnership for Research in Interfacial and Materials Engineering (IPRIME) for financial support.